\documentstyle[12pt]{article}
\begin{document}
\setcounter{page}{0}
\thispagestyle{empty}
\def\baselinestretch{1.5}
\begin{flushright}
{\bf IC/94/303}
\end{flushright}

\begin{center}
{\large\bf SIGNATURES OF VIRTUAL LSPs AT THE TEVATRON
 \\[0.5truein]}
{\large Amitava Datta\footnote{
Permanent address: Department of Physics, Jadavpur University ,
Calcutta - 700 032,India}\\}
International Centre for Theoretical Physics\\
34100 Trieste,ITALY\\
{\large Monoranjan Guchait\\}
Physics Department, Jadavpur University \\
Calcutta - 700 032, INDIA \\
and\\
{\large Surojit Chakravarti\\}
Maharaja Manindra Chandra College\\
20 Ram Kanta Bose Street\\
Calcutta - 700 003, INDIA
\vskip 10pt
{\bf ABSTRACT}
\end{center}
\sl
Relatively light sneutrinos which are experimentally allowed
and are not theoretically disfavoured may significantly affect
the currently popular  search  strategies  for  supersymmetric
particles by decaying dominantly into an invisible channel. In
certain cases the second lightest neutralino  may  also  decay
invisibly leading to two  extra  carriers  of  missing  energy
(in addition to the lightest supersymmetric particle (LSP) )
- the virtual LSPs  (VLSPs).  The  lighter  charginos  which  would be
produced in pairs  with  reasonably  large  cross-sections  at
TEVATRON energies,  decay  dominantly  into  the  hadronically
quiet  lepton  +  sneutrino  ($\not{E_T}$ )  modes  with  large
branching ratios leading to interesting unlike  sign  dilepton
events which are not swamped by the standard model  background.
The kinematical cuts required  to  eliminate  the  backgrounds
from WW, Drell-Yan and $\tau$ pair  production  are  discussed  in
detail. With 100 $pb^{-1}$ luminosity 10 - 35 background  free
events can be found in a large  region  of  the SUSY  parameter
space.

\rm
\vskip 10pt

\newpage
\def\baselinestretch{1.8}
\begin{center}
\bf\large I. ~ INTRODUCTION.
\end{center}
\normalsize\rm

Supersymmetry ( SUSY )
 \cite{1} provides an elegant solution of the
the notorious fine--tuning problem \cite{1}  which arises in the standard
model ( SM ). Moreover, this solution works only if the
supersymmetric partners of the known particles have masses $\sim$ 1
TeV. The lower end of this interesting mass spectrum
 is already accessible to some of the ongoing accelerators like
 the Fermilab Tevatron or LEP-- 100 at CERN. The planned accelerators
like the LEP-- 200 or the LHC at CERN can probe even higher mass scales.
The search for supersymmetry (SUSY)  is, therefore,
 a high priority programme.

 The strategy for hunting SUSY particles hinges on one crucial assumption:
there is a single, stable ( by virtue of
 a conserved quantum number (R-parity)), weakly interacting
particle, the so-called lightest supersymmetric particle (LSP). This particle
if produced easily escapes detection . It is further assumed that as a result
of the above  conservation law
 all other superparticles eventually decay into the LSP. The
LSP alone carries missing transverse energy ($\not{E_T}$) which
is traditionally regarded as the most distinctive signature of SUSY
particles.

The minimal supersymmetric extension of the standard model (MSSM)
contains four spin 1/2 neutral particles . These particles are
the super--partners of the photon, the Z boson and the two neutral
Higgs bosons. Linear combinations of these four states,the
 four neutralinos ($\widetilde{N_i}$, i=1-4),  are the physical
states. In the currently
favoured models, the lightest among them ($\widetilde{N_1}$) is
assumed to be the LSP \cite{1}.

 Recently it has been emphasised\cite{2,3}  that there may exist
SUSY particles which, though unstable, {\bf decay dominantly into
invisible channels}. This occurs if the sneutrinos ($\widetilde{\nu}$)
(the super--partners of the neutrinos) , though heavier than the LSP,
are lighter than the lighter chargino ($\widetilde{\chi_2}$)
and the second lightest neutralino
 ($\widetilde{N_2}$) and are much lighter than all other SUSY particles.
 As a cosequence, the {\bf invisible
two-body decay mode $\widetilde{\nu} \longrightarrow \nu \widetilde{N_1}$
 opens up} and completely dominates over others, being the only
kinematically allowed two-body decay of the sneutrinos. The other
neccesary condition for this scheme to work is that the
$\widetilde{N_1}$
has a substantial Zino component. This, however, is almost
always the case as long as the gluino (the super--partner
of the gluon) has a mass($m_{\widetilde{g}}$) in the range interesting for the
SUSY searches at the Tevatron \cite{4}. Moreover, in such cases the
 $\widetilde{N_2}$ which also has a dominant Zino component,
 decays primarily through the process
$\widetilde{N_2} \longrightarrow \nu \widetilde{\nu}$.
These two particles( $\widetilde{N_2}$ and $\widetilde{\nu}$ )
decaying primarily into invisible channels, hereafter
called {\bf virtual LSP(VLSP)'s}, may act as additional sources of $\not{E_T}$
and can significantly affect the strategies for SUSY search\cite{2}.

    Another important consequence of  relatively light sneutrinos is that
 $\widetilde{\chi_2}$ decays leptonically with a branching ratio( BR )
$\approx 1$.This occurs via the mode $\widetilde{\chi_2} \longrightarrow l
\widetilde{\nu}$ which is its only kinematically allowed two body decay.
In the following this deacy will play a crucial role,  with l = e or $\mu$.

So far  the SUSY search programmes  at hadron colliders \cite{4}are
primarily based on the  production of the strongly interacting particles
-- the squarks and the gluinos -- and their subsequent decays. The impact
of the VLSPs on these searches were considered in \cite{2}.
However as the  luminosity accumulates at the Tevatron, processes with
lower cross-sections may also become relevant in the SUSY hunt
programme \cite{5,6}.
Virtual LSP's can significantly affect these search strategies
as well. For example, in the conventional SUSY scenarios
 the process \cite{5,6}
$p \overline{p} \longrightarrow
 \widetilde{\chi_2} \widetilde{N_2} \longrightarrow 3l + \not{E_T}$ can probe
SUSY parameter ranges which are competitive with the reach of
LEP-II \cite{6}.
In models with virtual LSP's, thanks to the decay
$\widetilde{N_2} \longrightarrow \nu \widetilde{\nu}$, the final
state for the above process may consist only of a single lepton (from
$\widetilde{\chi_2} \longrightarrow l \widetilde{\nu}$) $+ \not{E_T}$. This
will obviously be overwhelmed by the backgrounds from
$W \longrightarrow l \nu$.

     The presence of virtual LSP's can, however, {\bf strengthen
the hadronically quiet $2l + \not{E_T}$ signal} coming from
$\widetilde{\chi_2}$-pair
production followed by the decay
 $\widetilde{\chi_2} \longrightarrow l \widetilde{\nu}$,
where l = e or $\mu$ \cite{2}.
In the conventional scenario this signal cannot compete with the
background, e.g. , from $WW$ production \cite{6}. The cross-section for the
latter process is about $10 pb$ \cite{7}, while the
$\widetilde{\chi_2} \widetilde{\chi_2}$
production cross-section is only $\sim 1 pb$. Since similar BR's are
involved in the signal and the background, the signal does not
seem to be promising \cite{6}. If, on the other hand,
virtual LSP's are present,
the situation changes drastically since
$B(\widetilde{\chi_2} \longrightarrow l \widetilde{\nu}) \approx 2/3$,
which is significantly larger
than $B(W \longrightarrow l \nu)$, where $l = e  or  \mu$. The
signal cross-section is then estimated to be $4/9 pb$, while the
background cross-section is approximately $2/5 pb$. In this paper we
analyse this hadronically quiet unlike sign dilepton signal more
quantitatively. We note that the sneutrinos being massive, the  lepton
spectrum in the decay   is expected to be softer than that of the
leptons from W decays. Exploiting this fact we can eliminate the lepton
background coming from $W-W$ pair production by imposing suitable kinematical
cuts.

The other important backgrounds that we have considered are the 1) Drell-
Yan pair production of $\tau^+-\tau^-$ and their subsequent leptonic decays,
2) $p \overline{p} \longrightarrow Z-Z$
with one Z decaying into an invisible mode and the other decaying leptonically,
3) $p \overline{p} \longrightarrow \ W-W $, one  W
decays leptonically into a  stable lepton (e or $\mu$), while the other decays
into a $\tau$ and a neutrino; the $\tau$ subsequently decays leptonically.
All these backgrounds and their removal are discussed in detail in
the next section, where we also present our results systematically.\\\

\begin{center}
\bf\large II. ~ Results and Discussions.
\end{center}
\normalsize\rm
We have generated the $\widetilde{\chi_2}-\widetilde{\chi_2}$
 events by a parton level Monte-Carlo
programme using the EHLQ-2 \cite{8} parton density functions. The formula
for chargino pair production cross-section is given, e.g , in \cite {9}. The
chargino pair productilon cross-section, thus computed, matches well with the
cross-sections given by others \cite{6}. Our choices of the SUSY
parameters are consistent with the constraints from LEP 100 \cite{10}.

Our signal for the chargino pair production is  hadronically
quiet opposite sign dilepton events (e or $\mu$). We show in the fig. 1 the
$P_T$
distribution of leptons (without any kinematic cuts) arising from the
decays of W's and charginos. We see that the leptons from W decay peak around
35 GeV as expected , whereas the leptons from the chargino decay are
considerably softer because this decay is associated with a heavy sneutrino.
We finally note that if events are selected  with {\bf both the leptons
having ${P_T} <$25 GeV} the
background is removed almost completely, without affecting signal very much.

     Drell-Yan $e^+ - e^-$ and ${\mu^+}-{\mu^-}$ events are expected
 to be back to back in the transverse plane and can be easily eliminated.
 However, ${\tau^+}-{\tau^-}$
events with both the $\tau$'s decaying leptonically
constitute a more serious background. These leptons essentially
move along the $\tau$ direction as a result of high boost. So these leptons
will also be nearly back to back in the transverse plane. Following the
CDF \cite{11} collaboration we have introduced an upper cut of 160$^\circ$ on
 $\phi$,
 the azimuthal angle between the lepotons in the transverse plane. Low
$P_T$
background leptons ,however,  survive this cut. We show in fig. 2 the
$P_T$ distribution of leptons
from the back ground and the signal with the cut on the azimuthal
angle as discussed above. The distribution clearly shows that the Drell-
Yan leptons which survive this cut
have $P_T$ peaking at low values. A cut of${P_T} > 10$ Gev
eliminates the remaining  background.

     In summary we demand that both the leptons should have $P_T$ within a
window of $ 10 GeV < {P_T} < 25 GeV$ and the azimuthal angle between the
leptons
$<$160$^\circ$. The requirement that both the leptons fall
within the $P_T$ window, very
efficiently reduces the background.

     We have also considered the   $\not{E_T}$
 distributiloins of chargino pair
events and WW events with the above $\phi$ and $\not{E_T}$ cut.
 Both the signal and the background
have large $\not{E_T}$. It turns out that  the conventional
 $\not{P_T}$ cut is
not particularly efficient in this case.

     We have also introduced CDF type cut \cite{12} on the invariant mass of
the lepton pair used in the context of the search for hadronically quiet
trilepton events and have required  $80 GeV < M_{ll} < 100 GeV$.
 We have used the isolation
cut between the two leptons similar to that used by the CDF group \cite{12}.
We require the isolation
$\Delta R= \sqrt{\Delta{\eta^2} + \Delta{\phi^2}}$
 between the leptons to be greater than 0.4 .
Apart from these cuts we have taken care of the CDF detector acceptance
for electrons and muons by introducing different pseudorapidity cuts as
described in \cite{12}.

     With all these cuts we have no background events from ZZ
production, with one Z
decaying invisibly and the other leptonically or from WW production,
with one W going
to e or$\mu$ and the other going to $\tau$ which then decays leptonically.

     We now discuss the signal events which survive all these cuts.
     We have calculated the signal cross-section with different values of
$m_{\widetilde{g}}$ and the $\mu$ parameter
for two values of tan$\beta$ (=2 and 10). All through we
have taken $m_{\widetilde{q}} $= $m_{\widetilde{g}}$ + 125 GeV.
 Table I shows the number of events surviving all
cuts with an integrated luminosity of 100 $ pb^{-1}$ for
tan$\beta$= 2 and Table II shows
the same for tan$\beta$ = 10.
For each set of parameters lighter chargino mass $m_{\widetilde{\chi_2}}$
has also been shown. We have taken the sneutrino mass to be  60 GeV.
The numbers of ee, e-$\mu$ and $\mu - \mu $ events are quoted differently
since the detector acceptances are different for e and $\mu$.
It is expected that with the main injector upgrade at the Tevatron the
integrated luminosity would  be as high as 1000$pb^{-1}$ by the end of
this century \cite{13}. It is quite clear from
the numbers given in the tables that
should this happen, SUSY search following the strategy chalked out in the paper
will be even more attractive.

  We see that even for low gluino mass, $m_{\widetilde{g}}$ = 150 GeV,
which is outside the domain of search
via the like sign dilepton mode in this scenario \cite{2} we expect
about 22 events for
$\mu$ = -100 GeV and 8 events for $\mu$ = - 200 GeV.

     For a higher  gluino mass ($m_{\widetilde{g}}$= 300 GeV),
which is likely to be beyond the reach of
the Tevatron by direct search following any conventional search strategy,
we still find about 10 - 35 background free events with $\mu$ positive and
 tan $\beta$ =
2. For tan $\beta$ = 10 only
$\mu$ = 200 - 300 GeV gives the above  number of  background free
events.

     It is obvious that the lepton $P_T$
spectrum from charginos sensetively depends on the
mass difference between the chargino and the sneutrino. For low mass
difference the spectrum becomes too soft to be distinguishable from the
Drell-Yan ${\tau^+}-{\tau^-}$ background. Again a large mass difference
makes the lepton spectrum
too hard to be distinguishable from the WW background. In the fig. 3 we
show for the purpose of illustration a plot of the
 number of surviving signal events with  100
$pb^{-1}$ luminosity
as a function of the sneutrino mass for a fixed mass of the chargino (=82
GeV). We see there is an optimal mass difference between the chargino and
the sneutrino where these cuts are most efficient. Thus for any mass
of the chargino we can actually probe the mass of the sneutrino
in a certain range.\\
      In this paper the masses of the of the SUSY particles were treated
as  free phenomenological parameters. It has , however, already been
argued that \cite{2}, the VLSP scenario can be accommodated in models based
on N = 1 Supergravity with common scalar and gaugino masses at a high
scale\cite {1}. For further details see \cite{14}.\\\

\begin{center}
\bf\large III. ~ Conclusions
\end{center}
\normalsize\rm

In conclusion we wish to reiterate that in the presence of the VLSP's
the hadronically quiet di- and tri- lepton events switch their roles.
As pointed out in ref 2,  the later signal which is at the focus of
current interests \cite{5,6,12} may be completely wiped out, while
the former looks indeed promising. In this paper we have quantified this claim
of ref 2 by calculating the signal at Tevatron energies. We have also
shown that the Standard Model background can be eliminated by a judicious
combination of  kinematical cuts. For a wide range of the SUSY parameter
space 10-35 back ground free events are estimated on the basis of an
integrated luminosity of 100 $pb^{-1}$. With the enhancement of the
 integrated luminosity due to a possible upgradation of the main injector
at the Tevatron, searches in this channel seems to be even more promising.

     At Tevatron energies the mass of
chargino and sneutrino that can be probed
by this method is limited by the available energy.
The strategy outlined in this paper can be extended in a straight forward
way to  LHC energies where  higher chargino and
sneutrino masses can be probed.
At LEP II SUSY signals for chargino masses in the range 80 - 90 GeV
 can also be looked for via the hadronically quiet
unlike sign dilepton events. The elimination of the backgrounds will
however be much easier due to  the characteristically
cleaner environment of a $e^+ - e^- $ machine.

{\bf Acknowledgements} : S.C thanks  the
Distributed Information Centre (DIC), Bose Institute, Calcutta for
computation facilities. A.D thanks the International Centre for Theoretical
Physics,Trieste  where part of this work was done , for hospitality and the
Department of Science and Technology, Government of India for a research
project under which part of this work was done.
The work of M.G. has been supported by the Council for
Scientific and Industrial Research, India.

\newpage
\centerline {\bf FIGURE CAPTIONS}

1)  $P_T$ distrinbution of leptons with no kinematical cuts. Continuous curve:
leptons from chargino decay. Dashed curve: leptons from W decay. SUSY
parameters: $m_{\widetilde{g}} = 200 GeV,
 tan \beta = 2, \mu = -200 GeV, m_{\widetilde{\chi_2}} = 82 GeV,
 m_{\widetilde{\nu}}$  = 60 GeV.

2)  $P_T$ distribution of leptons with $\phi < 160^\circ $ and
 other cuts (but no cut on $P_T$).Continuous curve:
leptons from chargino.
Dashed curve: leptons from Drell-Yan $\tau - \tau$
pair production followed by  leptonic  decays
of $\tau$'s. SUSY parameters: same as in figure 1.

3)  Variation of the number of signal events surving all cuts in 100 $pb^{-1}$
integrated luminosity as a function of the sneutrino mass .
SUSY parameters: $m_{\widetilde{g}}$ = 200 GeV, tan$\beta$ = 2,
$\mu$ = -200 GeV, $m_{\widetilde{\chi_2}}$ = 82 GeV.\\

\begin{table}
\caption{{\bf Number of dilepton events : tan$\beta$=2.0,
All masses in GeV.}}
\begin{center}
\begin{tabular}{||c|c|c|c|c|c|c||}
\hline \hline

$m_{\widetilde{g}}$ &$\mu$ &$m_{\widetilde{\chi_2}}$ &ee &$
\mu \mu$ &e-$\mu$ &Total no of events \\
\hline \hline
150 &-100 &71.6 &8 &4 &10 &22 \\
,, &-200 &67.5 &3 &1 &4 &8 \\
200 &-200 &82.5 &9 &4 &11 &24 \\
,, &-400 &77 &13 &6 &17 &36 \\
250 &-100 &92.7 &2 &1 &2 &5 \\
,, &-300 &95.2 &2 &1 &2 &5 \\
,, &500 &70.7 &8 &4 &11 &23 \\
,, &700 &74.7 &13 &6 &17 &36 \\
,, &300 &75.5 &13 &6 &17 &36 \\
,, &500 &86.8 &7 &2 &8 &17 \\
,, &700 &91.1 &4 &1 &4 &9 \\
\hline \hline
\end{tabular}
\end{center}
\end{table}
\vskip 15pt

\begin{table}
\caption{{\bf Number of dilepton events : tan$\beta$=10.0,
 All masses in GeV.}}
\begin{center}
\begin{tabular}{||c|c|c|c|c|c|c||}
\hline \hline

$m_{\widetilde{g}}$ &$\mu$ &$m_{\widetilde{\chi_2}}$ &ee &$
\mu \mu$ &e-$\mu$ &Total no of events \\
\hline \hline
200 &-300 &66.3 &1 &1 &2 &4 \\
250 &-400 &83.2 &10 &2 &4 &16 \\
,,  &300 &73.2 &11&5 &15 &31 \\
,, &-200 &67.5 &10 &3 &3 &16 \\
300 &-400 &99.2 &2 &0 &1 &3 \\
,, &200 &78.3 &12 &5 &14 &31 \\
,, &300 &88.4 &4 &2 &3 &9 \\
\hline \hline
\end{tabular}
\end{center}
\end{table}


\vskip 20pt
\thebibliography{19}

\bibitem{1} For  reviews see, for example, H.P.Nilles, Phys. Rep.
{\bf 110}, 1 (1984); P.Nath, R.Arnowitt and A.Chamseddine, Applied
N = 1 Supergravity , ICTP Series in Theo. Phys., Vol I, World
Scientific(1984);H. Haber and G. Kane,
Phys. Rep. {\bf 117},  75 (1985); S.P.Misra, Introduction to Supersymmetry
and Supergravity, Wiley Eastern, New Delhi (1992).
\bibitem{2} A.Datta, B.Mukhopadhyaya and M. Guchhait,PREPRINT MRI
-PHY-11/93, submitted for publication.
\bibitem{3} Some apsects of VLSPs in the context of
SUSY searches at hadron colliders
have  been considered by
H. Baer, C. Kao and X. Tata, Phys. Rev. {\bf D48},
 R2978 (1993); M. Barnett, J. Gunion and H. Haber, LBL preprint
LBL-34106 (1993).
\bibitem{4} CDF collaboration, F.abe et al., Phys. Rev.Lett.{\bf 69}
,3439(1992).
\bibitem{5} R. Arnowitt and P. Nath, Mod. Phys. Lett. {\bf A2},
 331(1987).
\bibitem{6} H. Baer and X. Tata, Phys. Rev. {\bf D47},  2739(1993).
\bibitem{7} See, for example, V. Barger and R. Phillips,
Collider Physics (Addison-wesley, 1987).
\bibitem{8} E.Eichten{\it et al}, Rev.Mod. Phys.{\bf 56},579(1984).
\bibitem{9} V.Barger{\ et al}, Phys.Lett.{\bf 131 B},372(1983).
\bibitem{10} H. Baer, M Drees, X. Tata, Phys. Rev. {\bf D41},
 3414(1990); G. Bhattacharyya, A. Datta, S. N. Ganguli and A. Raychaudhuri,
Phys. Rev. Lett. {\bf 64},  2870(1990);
A. Datta, M. Guchhait and A. Raychaudhuri, Z. Phys.
{\bf C54}, 513 (1992).
J. Ellis, G. Ridolfi and F. Zwirner, Phys. Lett.
{\bf B237}, 923(1990); M. Davier in Proc. Joint International Lepton-Photon
and Europhysics Conference in High Energy Physics, Geneva, 1992 (eds.
S. Hegarty {\it et al.}, World Scientific, 1992) p151.
\bibitem{11} CDF collaboration, F. Abe {\it et al.}, Phys.Rev.D
{\bf45},3921(1992).
\bibitem{12} CDF collaboration, Fermilab--conf--93/213--E CDF.
\bibitem{13} See, for example, J. Huth (CDF collaboration) in
Proc. XXVI International Conference on High Energy Physics,
Dallas, 1992 (ed J. R. Stanford, AIP Conference Proceedings no. 272)
p1273.
\bibitem{14} A.Datta, M.Drees and M.Guchait ,paper submitted to the
XXVII International Conf on High Energy Physics, Glasgow,1994,Contribution
Code: gls 0712 and Dortmund University pre-print (in preparation).
\end{document}